# Model Based Design of Super Schedulers Managing Catastrophic Scenario in Hard Real Time Systems

A. Christy Persya[1], T.R.Gopalakrishnan Nair[2]

[1]Research Associate, Real Time Systems Group, RIIC, Assistant Professor, BNMIT, India.
[2]Saudi Aramco Endowed Chair - Technology, Prince Mohammad University, Vice President, RIIC,
DS Institutions, India
[1]christypersya@gmail.com, [2]trgnair@gmail.com

**ABSTRACT**

The conventional design of real-time approaches depends heavily on the normal performance of systems and it often becomes incapacitated in dealing with catastrophic scenarios effectively. There are several investigations carried out to effectively tackle large scale catastrophe of a plant and how real-time systems must reorganize itself to respond optimally to changing scenarios to reduce catastrophe and aid human intervention. The study presented here is in this direction and the model accommodates catastrophe generated tasks while it tries to minimize the total number of deadline miss, by dynamically scheduling the unusual pattern of tasks. The problem is NP hard. We prove the methods for an optimal scheduling algorithm. We also derive a model to maintain the stability of the processes. Moreover, we study the problem of minimizing the number of processors required for scheduling with a set of periodic and sporadic hard real time tasks with primary/backup mechanism to achieve fault tolerance. EDF scheduling algorithms are used on each processor to manage scenario changes. Finally we present a simulation of super scheduler with small, medium and large real time tasks pattern for catastrophe management.

## 1. INTRODUCTION

Real time computing is an enabling technology for many current and future areas and has become increasingly pervasive. Real time systems often form a part of safety critical systems e.g. control systems for nuclear plants, aircrafts, or factories. The next generation real time systems must be designed to be dynamic, flexible and adaptable so that it can deal with certain unusual arrival of patterns of tasks arising out of abnormal situations like catastrophe. The real time system has a controlling system and it realizes a controlled environment. In this system, there could be unexpected or irregular events that must be attended immediately to minimize or avoid catastrophe. It calls for certain ability of the schedulers like the ability to dynamically alter the priorities of the tasks, change scenarios, and minimize catastrophic failures.

Catastrophe failure is a sudden, unexpected vital failure of the major parts of a system which can lead to high damage and even loss of human lives. A way need to be designed into in real time schedulers meant for catastrophe managing scenarios to contain or minimize the disastrous effects and initiate several recovery operations. One way to achieve this is through enabling intelligence over rigidity with which previous real-time systems were designed. It can help reducing the effects of these catastrophe failures during overloads and make the system restabilize using the super schedulers [1] having intelligent flexibility. For example, in nuclear power plants when there is a buildup of power output surge, and when an emergency shutdown was attempted, a more extreme spike in power output will occur, which will lead to a reactor vessel rupture and a series of explosions [10]. Or it can be in the case of an earthquake affecting the plant, or the aircraft in sudden vital system or influencing environment dysfunction. In such cases, since scenario change takes place from normal situations, there is a requirement for altered architecture and scheduling strategy to take actions and bring back either normalcy or minimize the effect of catastrophe. Here, a super scheduler can be called in to handle the scenario changes. It can handle both routine systems optimally and the catastrophe tasks satisfactorily by initiating the priority alter process and task re-assignment. Along with these, the new task introduction process is included which are the main features of the scheduler.

The super scheduler can also be used in a natural way to avoid catastrophe and to increase several catastrophe management phenomena. Here, we discuss why a catastrophic system is required and why the non catastrophic system cannot run by itself in these situations. How the priorities work with catastrophe system and how the dynamic context switching can takes place within that.

The paper is structured as follows. Section 2 discusses the scheduling theory and the problem definition. Section 3 states the literature survey of real time scheduling. Section 4 studies the super scheduler model and next section discusses the simulation study based on super scheduler. Section 6 concludes the paper.

## 1. SCHEDULING THEORY

The rationale of scheduling real time tasks is to have a precise schedule for all tasks such that all real time tasks can meet its deadline. Therefore during the occurrence of catastrophe the super scheduler can alter the priorities and reassign the tasks, rather than letting the complete scheduler into either failure or under performance as in the case recent failures in Fukoshima [11]. In this way, the system is allowed to maintain its stability by losing few optional real time tasks through reconfiguration, up gradation and disposal of low criticality.

Each task occurring in a real-time system has some timing properties. These timing properties should be considered when scheduling tasks on a real-time system.

The timing properties of a given task refer to the following [16].

• *Release time (or ready time $(r_j)$)*: Time at which the task is ready for processing.

• *Deadline $(d_j)$:* Time by which execution of the task should be completed.

• *Execution time $(e_j)$:* Time taken without interruption to complete the task, after the task is started.

• *Completion time $(C_j)$*: Maximum time taken to complete the task, after the task is started. This factor depends on the schedule.

•*Priority $(\zeta_j)$:* Relative urgency of the task.

• *Period $(P_j)$:* A periodic task $\tau_j$ is a task recurring at intervals of time. Period $P_j$ is the time interval between any two consecutive occurrences of task $\tau_j$.

A real-time application is normally composed of multiple tasks with different levels of criticality. Although missing deadlines is not desirable in a real-time system, soft real-time tasks could miss some deadlines and the system could still work correctly. However, missing some deadlines for soft real-time tasks will lead to paying penalties. On the other hand, hard real-time tasks cannot miss any deadline; otherwise, undesirable or catastrophe results will be produced in the system [16, 18].

We can formally define a real time system as follows.

Consider a system consisting of a set of tasks, $T=\{T_1,T_2,T_3,....T_n\}$, where the fairness of the schedule can be defined [4] as the sum of release time and execution time of a task should be less than the deadline of the task. To have a precise schedule each task $T_i$ from the set T should have its

$$ri + ei \leq di$$

The processor utilization factor [12] is given by the fraction of processor time spent in the execution of the task set

$$U = \sum_{i=1}^{n} \frac{ei}{pi}$$

### 1.1 PROBLEM DEFINITION

Consider the scenario, in which the system has arrival of hard real time tasks at higher inter arrival rate. The complex embedded real time systems have all hard deadlines with the relational dependency on other tasks. The problem is studied under overload conditions of tasks scheduled under hybrid scheduler. These overload conditions caused by the arrival of unusual patterns of most critical task will lead scenario change. Increasing the number of processors or reserving high end processors for the unusual arrival of catastrophe task will be costlier and occupy more space. The cost should be minimized while the system performance does not degrade. The problem is to dynamically change the priority of the currently running tasks with the newly arrived unusual catastrophe tasks. So, the currently running tasks are preempted by the catastrophe tasks. Since there is a serious emergency preemption, the stability of the processors should be well maintained. The stability of processors can be defined in terms of the minimum number of tasks miss its deadline. The problem is NP hard. For example if there are any unpredictable tasks like leakage of gas turbine in aircraft or burst in nuclear power plant, the system has to dynamically alter the priorities of the tasks and preempts the current execution.

Thus, the allocation of system resources must be well planned by the scheduler, so that all demands are met by the time required before deadline. This is usually done using a scheduler which implements a scheduling policy that determines how the resources of the system are allocated to the demands. We focus our attention to case the *scheduler* which reassigns the task to the processor in disjoint intervals and ensures that the system stability is maintained, where the problem is NP hard.

## 2. REAL TIME SCHEDULING

The set of tasks given as input should be ordered in such a way that they are executed with satisfactory constraints. The scheduling problems considered in this paper are characterized by a set of task $T=\{T_1,T_2,T_3,....T_n\}$ and a set of processors(Machines) $P=\{P_1,P_2,P_3,......,P_m\}$ on which the tasks execute.

A schedule is called feasible if every execution of the task meet its deadline. The parameter release time, execution time and deadline should be considered when scheduling tasks on real time system. The execution of tasks may or may not be preempted. Also, there may be precedence constraints, the order in which they execute. The fair scheduling will have few primary requirements to be satisfied.

The hard and soft real time tasks arrive in a queue to the scheduler. Scheduler allocates the arrived tasks to the processor and resources. Tasks are assigned using allocation algorithms like bin packing etc.

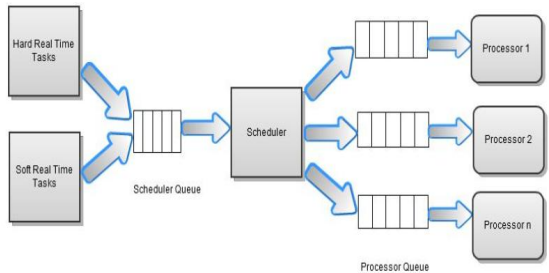

Fig 1. Block Diagram of a scheduler

Real time tasks can soft/ Hard or Periodic/ Aperiodic/ sporadic or fixed/ dynamic priority, preemptive/non preemptive or dependent/independent.

Many researchers have extensively worked on real-time scheduling algorithms. The *Rate Monotonic* (RM) scheduling algorithm is one of the most widely studied and used in practice [12, 14, 15, 16, 18]. It is a uniprocessor static-priority pre-emptive scheme. The RM algorithm and the *rate monotonic Deferred Server* (DS) scheduling algorithm are in the class of Pre-emptive Static-priority based algorithms [12, 14]. The *parametric dispatching* algorithm ([14, 20]) and the *predictive algorithm* ([14, 17]) are non-pre-emptive algorithms that attempt to provide high processor utilization while preserving task deadline guarantees and system schedulability.

The *Earliest Deadline First* (EDF) algorithm is a priority driven algorithm in which higher priority is assigned to the request that has earlier deadline, and a higher priority request always pre-empts a lower priority one [12, 14, 15, 16, 18, 19].

For scheduling a real time system, apart from release time, deadline and execution time it is required to know the importance of tasks as compared with other tasks. The feasible schedule ensures that all tasks meet its deadline with proper satisfactory constraints. An optimal scheduling algorithm is one which may fail to meet a deadline only when no other scheduling algorithm fails to meet. In order to prove the optimality of the scheduling algorithm, the feasibility conditions of the algorithm must be framed earlier. EDF is an optimal scheduling algorithm because there is no other dynamic scheduling algorithm which can successfully schedule all set of periodic tasks. Therefore, the optimal scheduling algorithm will be unique.

## 3. A SUPER SCHEDULER MODEL

In this section we provide a brief overview of the super scheduler. Before we begin to analyze this problem, we first express our assumptions as follows:

- A1. It has a set of independent periodic, sporadic and few optional tasks.
- A2. It has a perfect mapping of fault tolerance -- primary/backup model with identical multiprocessors.
- A3. The system periodically receives input from the external environment with a perfect timer.

Earlier [1] we discussed the theory of super scheduler. Super scheduler must be fault tolerant and more efficient [9] [10].The architecture of embedding hybrid scheduler into the super scheduler as shown in Figure 2. Until the arrival of catastrophe task $CT_i$, task T1 to Tn are feasibly scheduled by the hybrid scheduler.

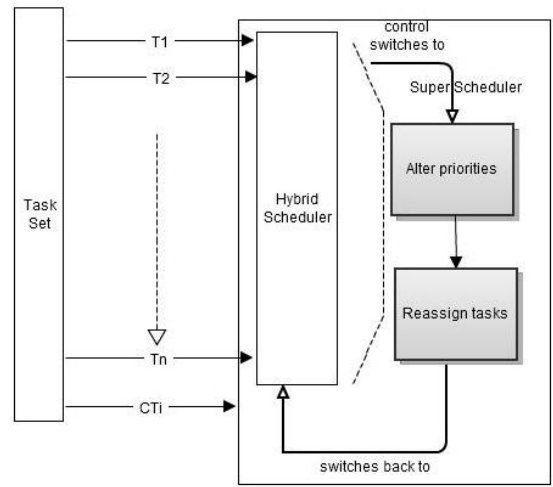

Fig 2. Embedding of hybrid into super scheduler

In major cycle, when the catastrophic task $CT_i$ arrives, the hybrid scheduler automatically switches its state to super scheduler. So the priority of the currently running high priority tasks can be altered with the arrived $CT_i$. The priority altered tasks are reassigned by the super scheduler. When the catastrophic task, $CT_i$ completes its process, then the context switching of altered tasks from the stack starts execution. Here, few low priority tasks and optional tasks cannot meet its deadline and some will be discarded, since they don't cause any major disaster. There is no additional execution time of the systems called up when the super scheduler is in command as well as in non-critical situations.

**Theorem 1:** Given a scheduler X in which the processor is tightly packed with the periodic tasks such that $T_i = \{c_i, d_i, p_i, r_i\}$ which are respectively compute, deadline, period and release time. The dispersion of $c_i$ and $r_i$ will determine the task and makes the scheduler unstable.

**Proof:**
By contradiction. Assuming two schedulers X, Y which are allotted with same set of tasks where, scheduler X is nicely packed with predetermined $c_i$ and $r_i$. The scheduler Y has the dispersion of $c_i$ and $r_i$. So, there is no predetermined schedule. The feasible schedule can be drawn from processor X and not for processor Y in all instances. It means that, with outstanding computations caused by arrival of unusual patterns of tasks, scheduler Y becomes unstable.

In accordance with theorem 1 the conclusion is that the scheduler X becomes unstable with the tasks dispersion of $c_i$ and $r_i$.

This state change can be described in the following model.

Here, the super scheduler ensures that the system remains in stable condition even under worst case situation and the behavior of the system is reliable.

As stated in [1] the stability condition of the system when scheduled by super scheduler after the catastrophe entry is equalized to 0.7.

Determinism and predictability are important characteristics of real time system. Most of the system is designed to withstand the worst case scenario. The system ensures that even in the worst case scenario, the behavior of the system is reliable.

No of Success task with catastrophe entry can be measured as

$$N_{success} = \frac{N - n_M}{N}$$

Where,

$n_M$ : No of least priority tasks missed its deadline

N : Total no of tasks super scheduled to the processor

No of Late tasks can be defined as

$$N_{late} = \sum_{i=1}^{n_b} late(t_i)$$

The task preemption here is completely different from the normal switching process. Here, along with the schedule in task reassignment the scenario which includes the environment, resource allocation vector and all factors which drives the system gets changed. So, when there is a sudden urge of scenario shift, the system stability has to be well maintained to avoid loss of lives or any danger.

The shift from scenario1 to N cannot be predicted in emergency conditions. Each scenario change causes a lot of changes to the system in all dimensions like new task releases, resource failures, deadlocks and more.

The new scenario change must be configured in few micro seconds and run time decisions have to be taken to avoid catastrophe. This decision making problem cannot be handled manually. It needs a next level of real time system. An approach towards this dimension is our super scheduler which is in the urge of designing the model which is NP-hard problem. The partial design phase of our super scheduler is analyzed in this paper.

The next property gives the guide of the super scheduler analysis.

**Property 1**

If task $CT_i$ makes a processor request at time t; the pre scheduled tasks from the set $\{T1, T2, .Tn\}$ are preempted at time t. Here, L1 set of tasks have started executing and L2 set of tasks are waiting for release time which is after t as shown below.

$$L1 @ t - t_{-1}$$
$$L2 @ t_1 - t$$

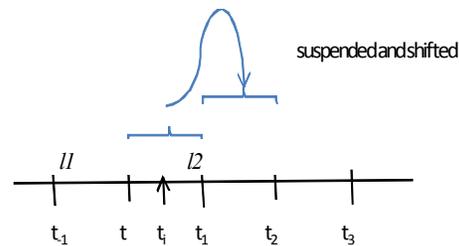

Suppose that a catastrophe task $CT_i$ arrives at time $t_i$, and there is no missed deadline before $t_i$. Let $t_{-1}$ be the latest time instant before t at which some job with a deadline after t executes. Suppose that $CT_k$ jobs execute $t_n$ the interval (t, t1]. We call these jobs $CT_1$,

$CT_2$, $CT_k$ and order them in increasing order of their deadlines. Because the processor remains busy with $CT_k$ jobs in $(t,t_1]$ executing the catastrophe task by altering their priorities with the high priority task $T_k$ and $T_{k+1}$ miss its deadline at time $t_1$, we must have

$$\sum_{i=0}^{L} \varrho_i > t_1 - t$$

Let the number of job-releases and job-completions during the time interval $(t-t_1)$ be $L$ and $t_i$ be the time instant at which $CT_i$ arrives. Then $\Delta_i$ denotes the total density of the $CT_i$ jobs.

Count of scheduled CS

$$CS(t - t_{-1}) + CS(t_1 - t) + CS(t_i) = N$$

Count of Executed CE

$$CE(t - t_{-1}) + CE(t_1 - t) + CE(t_i) = N - n_M$$

Because $CS(t_i) = CE(t_i)$

At time instant $t_i$,
$$CE(t - t_{-1}) + CS(t_1 - t) + CE(t_i) = 1$$

When there exist a feasible schedule of a set J of jobs with arbitrary release times and deadlines on a processor, the unusual arrival of a catastrophic task $CT_j$, can also be scheduled by suspending few low priority of tasks.

This statement can be proved with supporting statements given by Jane [4]
1. There will be no time to do online validation i.e., acceptance test, when the application creates a new hard task i.e., unpredictable catastrophe task.
2. In typical real time system, the system must maintain information on all existing hard real time tasks, including the no of such tasks. The number of tasks may change as tasks are added and deleted while the system executes.
3. Almost every real time system is required to respond to external events which occur at random instants of time.
4. During a transient overload when it is not possible to complete all the jobs in time, choose to discard optional jobs, so that the mandatory jobs can complete in time.
5. It is better not to execute than to execute late.

The initial real time system will be loaded with the possible and planned real time tasks. The planned real time tasks will be nicely scheduled by the optimal scheduler in a periodic style. Few possible tasks arrive at the random rate with urgent priority. The planned optional tasks can be discarded and the newly added tasks can get the chance to execute to avoid catastrophe by reprioritizing.

### 4.1 Task Set Generation

The task sets used to apply experiments are generated randomly according to the following criteria:

- The number of tasks is greater than 4.
- Deadlines are equal to periods
- The task set is feasible by EDF scheduler.

### 4.2 Results

Here, the number of tasks missed its deadline by the arrival of unusual pattern of catastrophe task caused by the super scheduler is analyzed. The results are observed as an average number of miss caused in 25 time units. The number of deadline miss caused by catastrophe task is always higher in normal EDF scheduler than our super scheduler.

The function formed with the dividend interpolation theory for the super scheduler execution is

$Y=-0.0001X^4+0.0044X^3-0.0558X^2+0.4663X-0.2284$

### 4. SIMULATION STUDIES

According to the design of the super scheduler for concurrent real time tasks at catastrophe it will be better, so it is necessary to measure their efficiency through an approach based on their timing constraints.

The medium system as shown in table 1 with an catastrophe entry is scheduled with [2], [3], [4] as

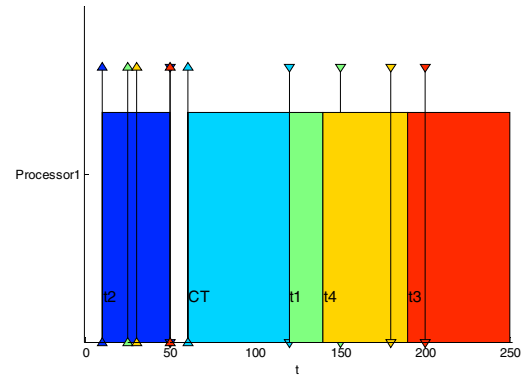

Fig 3.Medium System

| Task | Proc Time | Release Time | Deadline |
|------|-----------|--------------|----------|
| T1 | 20 | 25 | 150 |
| T2 | 40 | 10 | 50 |
| T3 | 60 | 50 | 200 |
| T4 | 50 | 30 | 180 |
| CT | 60 | 60 | 120 |

Table 1. Medium System

Here, because of catastrophe entry t3 miss its deadline. The stability of the system is maintained with the success rate of 0.7 as shown in (1).

The large system is considered with 10 tasks as shown in table 2.

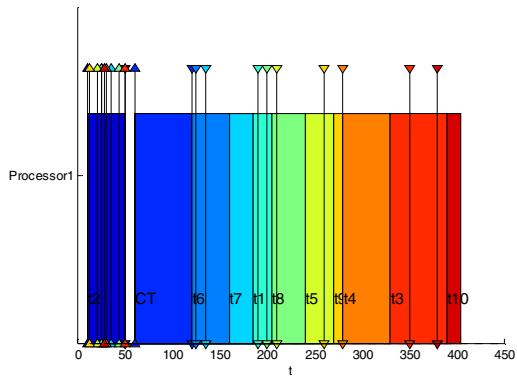

Fig 4. Large system

| Task | Proc Time | Release Time | Deadline |
|------|-----------|--------------|----------|
| T1 | 20 | 25 | 190 |
| T2 | 40 | 10 | 50 |
| T3 | 60 | 50 | 350 |
| T4 | 50 | 30 | 280 |
| T5 | 30 | 20 | 210 |
| T6 | 40 | 25 | 125 |
| T7 | 25 | 35 | 135 |
| T8 | 35 | 43 | 200 |
| T9 | 10 | 12 | 260 |
| T10 | 15 | 28 | 380 |
| CT | 60 | 60 | 120 |

Table 2. Large System

In large system, when 10 tasks are feasibly scheduled under hybrid scheduler and when CT enters, 2 tasks miss its deadline. So the stability of the system is maintained with the success rate greater than 0.7 as shown in equation (1).

In the same way, even for the large system the stability condition will be satisfied according to equation (1).

## 5. CONCLUSION

In this paper we provided a systematic way for realizing a scheduler in terms of design, simulation and analysis for an ideal catastrophe management. We estimated that the stability of the system will be maintained even with the sudden arrival of catastrophic task with the system having large set of hard deadline tasks and soft deadline tasks.

As a part of the future work, it can be improved to contain e the catastrophe management with cognitive control on the super scheduler.

**References**
[1] T.R.Gopalakrishnan Nair; A. Christy Persya, "Critical Task Re-assignment Under Hybrid Scheduling Approach in Multiprocessor Real-time Systems," Parallel and Distributed Computing and Systems, 2011. PDCS 2011, USA. 23rd IASTED International Conference on, vol., no., pp.130-137, 14-16 Dec. 2011doi: 10.2316/P.2011.757-071
[2] P. Šůcha, M. Kutil, M. Sojka, Z. Hanzálek. TORSCHE Scheduling Toolbox for Matlab. In IEEE International Symposium on Computer-Aided Control Systems Design. Munich, Germany: 2006.
[3] Rao, M.V.P. Shet, K.C. ; Balakrishna, R. ; Roopa, K.," Development of Scheduler for Real Time and Embedded System Domain", in Advanced Information Networking and Applications - Workshops, 2008. AINAW 2008. 22nd International Conference, 25-28 March 2008, 1- 6.
[4] Jane W.S.Liu, Real-Time Systems, 2nd Edition. Pearson Education, 2005.
[5] Shmueli, E.; Feitelson, D.G.; , "On Simulation and Design of Parallel-Systems Schedulers: Are We Doing the Right Thing?," Parallel and Distributed Systems, IEEE Transactions on , vol.20, no.7, pp.983-996, July 2009
doi: 10.1109/TPDS.2008.152
[6] Insop Song; Sehjeong Kim; Karray, F.; , "A Real-Time Scheduler Design for a Class of Embedded Systems," Mechatronics, IEEE/ASME Transactions on , vol.13, no.1, pp.36-45, Feb. 2008
doi: 10.1109/TMECH.2007.915061
[7] Mooney, V.J., III; De Micheli, G.; "Real time analysis and priority scheduler generation for hardware-software systems with a synthesized run-time system," Computer-Aided Design, 1997. Digest of Technical Papers., 1997 IEEE/ACM International Conference on, vol., no., pp.605-612, 9-13 Nov 1997doi: 10.1109/ICCAD.1997.643601
[8] Golatowski, F.; Hildebrandt, J.; Blumenthal, J.; Timmermann, D.; "Framework for validation, test and analysis of real-time scheduling algorithms and scheduler implementations," Rapid System Prototyping, 2002. Proceedings. 13th IEEE

International Workshop on, vol., no., pp. 146- 152, 2002doi: 10.1109/IWRSP.2002.1029750

[9] Echague, J.; Ripoll, I.; Crespo, A.; "Hard real-time preemptively scheduling with high context switch cost," Real-Time Systems, 1995. Proceedings. Seventh Euromicro Workshop on , vol., no., pp.184-190, 14-16 Jun 1995

doi: 10.1109/EMWRTS.1995.514310

[10]http://www.world-nuclear.org/info/Chernobyl/inf07.html

[11]http://www.world-nuclear.org/info/fukushima_accident_inf129.html

[12] C. M. Krishna, Kang G. Shin, Real-Time Systems, McGraw-Hill, 1997.

[13]. Technical Report No. 2005-499 Scheduling Algorithms for Real-Time Systems Arezou Mohammadi and Selim G. AklSchool of Computing

[14] K. Frazer, "Real-time Operating System SchedulingAlgorithms,"1997,

http://home.earthlink.net/krfrazer/Scheduler.pdf.

[15] W. A. Halang and A. D. Stoyenko, "Real Time Computing," NATO ASI Series, Series F: Computer and Systems Sciences, Springer-Verlag, Vol. 127, 1994.

[16] M. Joseph, "Real-time Systems: Specification, Verification and Analysis," Prentice Hall, 1996.

[17] H. Singh, "Scheduling Techniques for real-time applications consisting of periodic task sets," Proceedings of the IEEE Workshop on Real-Time Applications, pp. 12-15 , July 21-22, 1994.

[18] J. A. Stankovic and K. Ramamritham, "Tutorial on Hard Real-Time Systems,"

IEEE Computer Society Press, 1988.

[19] J. A. Stankovic, M. Spuri, K. Ramamritham, and G. C. Buttazzo, "Deadline Scheduling for Real-Time Systems, EDF and related algorithms," KluwerAcademia Publishers, 1998.

[20] R. Gerber, S. Hong and M. Saksena, "Guaranteeing Real-Time Requirements with Resource-Based Calibration of Periodic Processes," IEEE Transactions on Software Engineering, No. 7, Vol. 21, July, 1995.